\documentclass[journal]{vgtc}                     

\onlineid{0}

\vgtccategory{Research}

\title{Replication Studies: Not Just a Copy}

\author{%
  \authororcid{Yiheng Liang}{0009-0006-8029-2921},
  \authororcid{Kim Marriott}{0000-0002-9813-0377}, and 
  \authororcid{Helen C. Purchase}{0000-0001-6994-4446}
}

\authorfooter{
  \item
    Yiheng Liang, Kim Marriott and Helen C. Purchase are with Monash University. E-mail: \{Yiheng.Liang, Kim.Marriott, Helen.Purchase\}@monash.edu.
}

\abstract{Replication studies revisit previous experimental findings for multiple purposes, including assessing reliability, exploring generalisability, and evaluating research methods. Such studies entail design decisions that a single replication label cannot fully capture, making their designs difficult to describe and compare. We present REPVIS2, a validated design space that describes how a replication study relates to a reference study across eight practical dimensions, each coded as identical, similar, or different. We refined our initial design space, REPVIS1, by applying it to a corpus of replication studies and validated REPVIS2 against a separate corpus. We then characterised 86 replication studies from 51 papers in visualisation through paired reading of the replication and reference reports. Studies often retained the task while changing the procedure, interface, environment, participant population, evidence, or analysis. Additions beyond the replication core were also common. REPVIS2 makes replication design explicit for characterisation, reporting, and planning. An interactive visualisation and supplementary materials are available at \url{https://replication-study.github.io/}.}


\teaser{
  \centering
\resizebox{\linewidth}{!}{%
\begin{tikzpicture}[x=7.25in,y=3.55in]
\path[use as bounding box] (0,0) rectangle (1,1);
\node[anchor=south west,inner sep=0pt] at (0,0) {\includegraphics[width=7.25in]{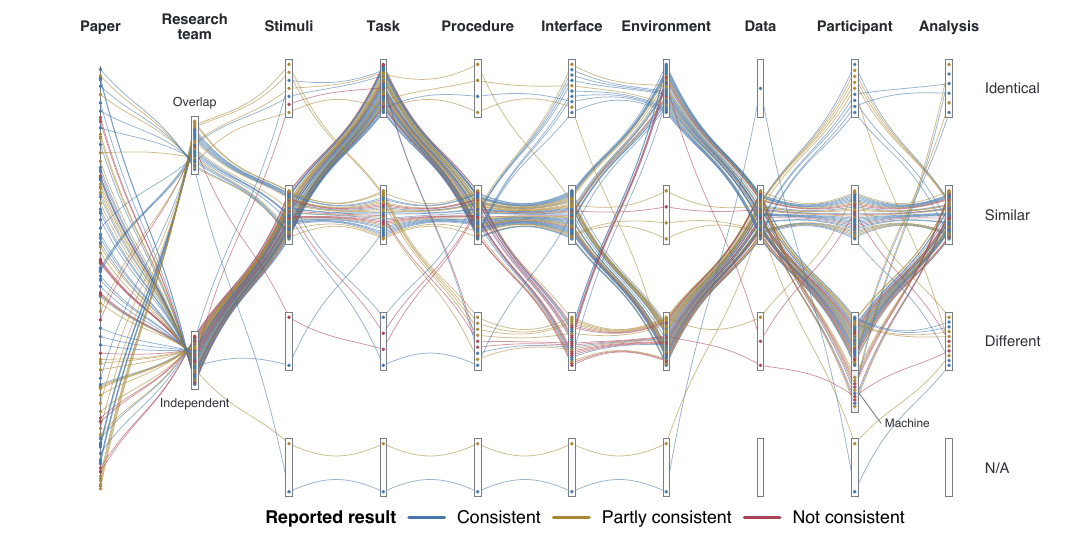}};
\node[anchor=east,inner sep=0pt,outer sep=0pt,text=black,font=\sffamily\fontsize{4.15}{4.5}\selectfont] at (0.0862797465909496,0.869582453494544) {\cite{repvisCorpus01} Guimbretière07};
\node[anchor=east,inner sep=0pt,outer sep=0pt,text=black,font=\sffamily\fontsize{4.15}{4.5}\selectfont] at (0.0862797465909496,0.853910844976892) {\cite{repvisCorpus02} Heer10};
\node[anchor=east,inner sep=0pt,outer sep=0pt,text=black,font=\sffamily\fontsize{4.15}{4.5}\selectfont] at (0.0862797465909496,0.838239236459241) {\cite{repvisCorpus03} Kong10};
\node[anchor=east,inner sep=0pt,outer sep=0pt,text=black,font=\sffamily\fontsize{4.15}{4.5}\selectfont] at (0.0862797465909496,0.82256762794159) {\cite{repvisCorpus04} Kosara10};
\node[anchor=east,inner sep=0pt,outer sep=0pt,text=black,font=\sffamily\fontsize{4.15}{4.5}\selectfont] at (0.0862797465909496,0.806896019423938) {\cite{repvisCorpus05} Hullman11};
\node[anchor=east,inner sep=0pt,outer sep=0pt,text=black,font=\sffamily\fontsize{4.15}{4.5}\selectfont] at (0.0862797465909496,0.791224410906287) {\cite{repvisCorpus06} Kim12};
\node[anchor=east,inner sep=0pt,outer sep=0pt,text=black,font=\sffamily\fontsize{4.15}{4.5}\selectfont] at (0.0862797465909496,0.775552802388636) {\cite{repvisCorpus07} Micallef12};
\node[anchor=east,inner sep=0pt,outer sep=0pt,text=black,font=\sffamily\fontsize{4.15}{4.5}\selectfont] at (0.0862797465909496,0.759881193870984) {\cite{repvisCorpus08} Ottley12};
\node[anchor=east,inner sep=0pt,outer sep=0pt,text=black,font=\sffamily\fontsize{4.15}{4.5}\selectfont] at (0.0862797465909496,0.744209585353333) {\cite{repvisCorpus09} Talbot12};
\node[anchor=east,inner sep=0pt,outer sep=0pt,text=black,font=\sffamily\fontsize{4.15}{4.5}\selectfont] at (0.0862797465909496,0.728537976835682) {\cite{repvisCorpus10} Harrison13};
\node[anchor=east,inner sep=0pt,outer sep=0pt,text=black,font=\sffamily\fontsize{4.15}{4.5}\selectfont] at (0.0862797465909496,0.71286636831803) {\cite{repvisCorpus11} Jakobsen13};
\node[anchor=east,inner sep=0pt,outer sep=0pt,text=black,font=\sffamily\fontsize{4.15}{4.5}\selectfont] at (0.0862797465909496,0.697194759800379) {\cite{repvisCorpus12} Ziemkiewicz13};
\node[anchor=east,inner sep=0pt,outer sep=0pt,text=black,font=\sffamily\fontsize{4.15}{4.5}\selectfont] at (0.0862797465909496,0.681523151282727) {\cite{repvisCorpus13} Harrison14};
\node[anchor=east,inner sep=0pt,outer sep=0pt,text=black,font=\sffamily\fontsize{4.15}{4.5}\selectfont] at (0.0862797465909496,0.665851542765076) {\cite{repvisCorpus14} Talbot14};
\node[anchor=east,inner sep=0pt,outer sep=0pt,text=black,font=\sffamily\fontsize{4.15}{4.5}\selectfont] at (0.0862797465909496,0.650179934247425) {\cite{repvisCorpus15} Le14};
\node[anchor=east,inner sep=0pt,outer sep=0pt,text=black,font=\sffamily\fontsize{4.15}{4.5}\selectfont] at (0.0862797465909496,0.634508325729773) {\cite{repvisCorpus16} Okoe15};
\node[anchor=east,inner sep=0pt,outer sep=0pt,text=black,font=\sffamily\fontsize{4.15}{4.5}\selectfont] at (0.0862797465909496,0.618836717212122) {\cite{repvisCorpus17} Kay15};
\node[anchor=east,inner sep=0pt,outer sep=0pt,text=black,font=\sffamily\fontsize{4.15}{4.5}\selectfont] at (0.0862797465909496,0.603165108694471) {\cite{repvisCorpus19} Ottley15};
\node[anchor=east,inner sep=0pt,outer sep=0pt,text=black,font=\sffamily\fontsize{4.15}{4.5}\selectfont] at (0.0862797465909496,0.587493500176819) {\cite{repvisCorpus48} Jansen15};
\node[anchor=east,inner sep=0pt,outer sep=0pt,text=black,font=\sffamily\fontsize{4.15}{4.5}\selectfont] at (0.0862797465909496,0.571821891659168) {\cite{repvisCorpus18} Kosara16};
\node[anchor=east,inner sep=0pt,outer sep=0pt,text=black,font=\sffamily\fontsize{4.15}{4.5}\selectfont] at (0.0862797465909496,0.556150283141517) {\cite{repvisCorpus20} Skau16};
\node[anchor=east,inner sep=0pt,outer sep=0pt,text=black,font=\sffamily\fontsize{4.15}{4.5}\selectfont] at (0.0862797465909496,0.540478674623865) {\cite{repvisCorpus21} Dimara16};
\node[anchor=east,inner sep=0pt,outer sep=0pt,text=black,font=\sffamily\fontsize{4.15}{4.5}\selectfont] at (0.0862797465909496,0.524807066106214) {\cite{repvisCorpus22} Feng16};
\node[anchor=east,inner sep=0pt,outer sep=0pt,text=black,font=\sffamily\fontsize{4.15}{4.5}\selectfont] at (0.0862797465909496,0.509135457588563) {\cite{repvisCorpus23} Dragicevic17};
\node[anchor=east,inner sep=0pt,outer sep=0pt,text=black,font=\sffamily\fontsize{4.15}{4.5}\selectfont] at (0.0862797465909496,0.493463849070911) {\cite{repvisCorpus25} Szafir17};
\node[anchor=east,inner sep=0pt,outer sep=0pt,text=black,font=\sffamily\fontsize{4.15}{4.5}\selectfont] at (0.0862797465909496,0.47779224055326) {\cite{repvisCorpus26} Burlinson17};
\node[anchor=east,inner sep=0pt,outer sep=0pt,text=black,font=\sffamily\fontsize{4.15}{4.5}\selectfont] at (0.0862797465909496,0.462120632035608) {\cite{repvisCorpus24} Reda18};
\node[anchor=east,inner sep=0pt,outer sep=0pt,text=black,font=\sffamily\fontsize{4.15}{4.5}\selectfont] at (0.0862797465909496,0.446449023517957) {\cite{repvisCorpus27} Haehn18};
\node[anchor=east,inner sep=0pt,outer sep=0pt,text=black,font=\sffamily\fontsize{4.15}{4.5}\selectfont] at (0.0862797465909496,0.430777415000306) {\cite{repvisCorpus28} Hu19};
\node[anchor=east,inner sep=0pt,outer sep=0pt,text=black,font=\sffamily\fontsize{4.15}{4.5}\selectfont] at (0.0862797465909496,0.415105806482654) {\cite{repvisCorpus29} Yang19};
\node[anchor=east,inner sep=0pt,outer sep=0pt,text=black,font=\sffamily\fontsize{4.15}{4.5}\selectfont] at (0.0862797465909496,0.399434197965003) {\cite{repvisCorpus32} Weiß20};
\node[anchor=east,inner sep=0pt,outer sep=0pt,text=black,font=\sffamily\fontsize{4.15}{4.5}\selectfont] at (0.0862797465909496,0.383762589447352) {\cite{repvisCorpus46} Heyer20};
\node[anchor=east,inner sep=0pt,outer sep=0pt,text=black,font=\sffamily\fontsize{4.15}{4.5}\selectfont] at (0.0862797465909496,0.3680909809297) {\cite{repvisCorpus31} Wu21};
\node[anchor=east,inner sep=0pt,outer sep=0pt,text=black,font=\sffamily\fontsize{4.15}{4.5}\selectfont] at (0.0862797465909496,0.352419372412049) {\cite{repvisCorpus34} Wesslen21};
\node[anchor=east,inner sep=0pt,outer sep=0pt,text=black,font=\sffamily\fontsize{4.15}{4.5}\selectfont] at (0.0862797465909496,0.336747763894398) {\cite{repvisCorpus35} McColeman21};
\node[anchor=east,inner sep=0pt,outer sep=0pt,text=black,font=\sffamily\fontsize{4.15}{4.5}\selectfont] at (0.0862797465909496,0.321076155376746) {\cite{repvisCorpus30} Syeda22};
\node[anchor=east,inner sep=0pt,outer sep=0pt,text=black,font=\sffamily\fontsize{4.15}{4.5}\selectfont] at (0.0862797465909496,0.305404546859095) {\cite{repvisCorpus33} Panavas22};
\node[anchor=east,inner sep=0pt,outer sep=0pt,text=black,font=\sffamily\fontsize{4.15}{4.5}\selectfont] at (0.0862797465909496,0.289732938341444) {\cite{repvisCorpus47} Sharif22};
\node[anchor=east,inner sep=0pt,outer sep=0pt,text=black,font=\sffamily\fontsize{4.15}{4.5}\selectfont] at (0.0862797465909496,0.274061329823792) {\cite{repvisCorpus40} Syeda23};
\node[anchor=east,inner sep=0pt,outer sep=0pt,text=black,font=\sffamily\fontsize{4.15}{4.5}\selectfont] at (0.0862797465909496,0.258389721306141) {\cite{repvisCorpus36} Davis24};
\node[anchor=east,inner sep=0pt,outer sep=0pt,text=black,font=\sffamily\fontsize{4.15}{4.5}\selectfont] at (0.0862797465909496,0.24271811278849) {\cite{repvisCorpus37} While24};
\node[anchor=east,inner sep=0pt,outer sep=0pt,text=black,font=\sffamily\fontsize{4.15}{4.5}\selectfont] at (0.0862797465909496,0.227046504270838) {\cite{repvisCorpus38} Matzen24};
\node[anchor=east,inner sep=0pt,outer sep=0pt,text=black,font=\sffamily\fontsize{4.15}{4.5}\selectfont] at (0.0862797465909496,0.211374895753187) {\cite{repvisCorpus45} vanGemert24};
\node[anchor=east,inner sep=0pt,outer sep=0pt,text=black,font=\sffamily\fontsize{4.15}{4.5}\selectfont] at (0.0862797465909496,0.195703287235536) {\cite{repvisCorpus39} Creamer25};
\node[anchor=east,inner sep=0pt,outer sep=0pt,text=black,font=\sffamily\fontsize{4.15}{4.5}\selectfont] at (0.0862797465909496,0.180031678717884) {\cite{repvisCorpus41} Long25};
\node[anchor=east,inner sep=0pt,outer sep=0pt,text=black,font=\sffamily\fontsize{4.15}{4.5}\selectfont] at (0.0862797465909496,0.164360070200233) {\cite{repvisCorpus42} Cutler25};
\node[anchor=east,inner sep=0pt,outer sep=0pt,text=black,font=\sffamily\fontsize{4.15}{4.5}\selectfont] at (0.0862797465909496,0.148688461682582) {\cite{repvisCorpus43} Khalaila25};
\node[anchor=east,inner sep=0pt,outer sep=0pt,text=black,font=\sffamily\fontsize{4.15}{4.5}\selectfont] at (0.0862797465909496,0.13301685316493) {\cite{repvisCorpus44} Wang25};
\node[anchor=east,inner sep=0pt,outer sep=0pt,text=black,font=\sffamily\fontsize{4.15}{4.5}\selectfont] at (0.0862797465909496,0.117345244647279) {\cite{repvisCorpus49} DiBartolomeo26};
\node[anchor=east,inner sep=0pt,outer sep=0pt,text=black,font=\sffamily\fontsize{4.15}{4.5}\selectfont] at (0.0862797465909496,0.101673636129627) {\cite{repvisCorpus50} Satkunarajan26};
\node[anchor=east,inner sep=0pt,outer sep=0pt,text=black,font=\sffamily\fontsize{4.15}{4.5}\selectfont] at (0.0862797465909496,0.0860020276119761) {\cite{repvisCorpus51} D'Adamo26};
\end{tikzpicture}%
}%

  \caption{%
  	Profiles of the 86 replication studies. Lines connect studies, research team, and the eight design dimensions; color encodes the reported replication result.%
  }
  \label{fig:replication-profiles}
}

\graphicspath{{figs/}{figures/}{pictures/}{images/}{./}} 

\usepackage{tabu}   
\usepackage{booktabs}                  
\usepackage{lipsum}                    
\usepackage{mwe}                       
\usepackage{ccicons}                   

\usepackage{mathptmx}                  

\usepackage{svg}
\usepackage{booktabs}
\usepackage{tabularx}
\usepackage{array}
\usepackage{amsmath}
\usepackage[normalem]{ulem}

\begin{document}


\firstsection{Introduction}

\maketitle

Visualisation research relies heavily on empirical studies to explain how people interpret visual representations, interact with systems, and make decisions from data~\cite{lam2012_seven_scenarios,isenberg2013systematic}. The findings of these studies often inform later theories, systems, and design guidance. Replication is therefore important for examining whether published findings remain reliable, where they generalise, and under what cases they change~\cite{nosek2020_replication,kosara2018_replication_crisis}. Without this process, the field risks building new work on results whose scope and robustness remain uncertain.

While replication might be rarer than the research community might like, it is not non-existent. Researchers have revisited classic graphical perception results, re-evaluated visualisation techniques and repeated studies with new populations and settings~\cite{repvisCorpus02,repvisCorpus14,repvisCorpus33}. However, these studies are distributed across different topics and research contexts. They also use inconsistent terminology and are not always explicit about their relationship to earlier work~\cite{plesser2018_terminology,hornbaek2014_once_enough}. As a result, the field has accumulated replication studies without yet developing an easily understood overall account of how those studies were designed, why they were conducted, and what they found.

A replication is more than simply a copy of its original reference study. It is a constrained study design in which some choices are preserved, others are adapted, and new components may be introduced. Existing terms such as \textit{strict replication} and \textit{conceptual replication}, together with strategies such as triangulation, generalisation, and specification, provide useful high-level descriptions~\cite{plesser2018_terminology,repvis1}. However, assigning one label to an entire study compresses many consequential design decisions into a single category. Such choices can matter to the resulting evidence; for example, the effectiveness of visual encodings can vary with the task and the distribution of the displayed data~\cite{kim2018assessing}. Studies carrying the same label may change different parts of an experiment, while studies described using different labels may remain comparable in many practical respects. What is needed is a way to make the structure of a replication design explicit.

This problem is not unique to visualisation. Other areas of technology-oriented human-subject research, such as broader Human-Computer Interaction (HCI), also need to explain how a replication study preserves, varies, or extends an earlier design~\cite{hornbaek2014_once_enough}. A design space account of these relationships may therefore inform replication research beyond visualisation.

Our initial design space, REPVIS~\cite{repvis1}, was published as a EuroVis short paper. We refer to it here as REPVIS1 to distinguish it from the revised version. It represented replication through multiple design dimensions, making the correspondence between a replication study and its reference study explicit. However, REPVIS1 was a conceptual proposal developed from prior terminology, advocacy, and a limited set of examples. It had not been systematically applied to a broader corpus, leaving open whether its structure could represent the diversity of replication designs found in practice.

This work addresses that gap through three research questions:

\noindent\textbf{RQ1} How can the design of a visualisation replication study be systematically represented in relation to its reference study, without reducing it to a single replication label?

\noindent\textbf{RQ2} What patterns characterise accumulated visualisation replication practice, including how studies are designed, why they are conducted, and what outcomes they report?

\noindent\textbf{RQ3} What do the resulting design space and the observed practices together suggest for planning and reporting future replication research?

To answer these questions, we tested our initial REPVIS1 on a set of existing replication studies (the 'development corpus'), and, as a result, refined it to produce the extended version (REPVIS2), which we evaluated using a 'validation corpus'. We then used REPVIS2 to analyse all the studies in both corpuses (51 papers and 86 replication studies, since a paper can contain multiple studies). Our contributions are:

\begin{itemize}
    \item a validated replication design space for systematically representing how a replication study relates to its reference study;

    \item a corpus of 86 replication studies from 51 papers in visualisation, with a characterisation of their designs, motivations, reported results, and recurring design choices;

    \item corpus-grounded implications for making replication relationships more explicit and for planning future visualisation replication research.
\end{itemize}

\section{Related Work}
\label{sec:related-work}

Throughout this paper, we use replication as an umbrella term for studies that revisit the same or a closely corresponding research question and compare their findings to an original reference study.

\subsection{Replication in Visualisation and Related Fields}

Replication is concerned with what earlier findings allow us to conclude, not simply whether an experimental procedure can be repeated. Nosek and Errington frame replication in terms of whether its possible outcomes would provide evidence about a previous claim~\cite{nosek2020_replication}. However, the meanings of replication vary across research traditions~\cite{plesser2018_terminology}. These differences make it important to examine the relationship between studies rather than infer it from terminology alone.

In visualisation, calls for replication have addressed both research validity and the conditions needed to revisit published work. Kosara and Haroz identify threats to study validity and discuss practices such as sharing materials, preregistration, and publication models that recognise the value of replication~\cite{kosara2018_replication_crisis}. This work connects replication to everyday research and publication practices, rather than treating it only as a response to disputed findings.

Published studies also demonstrate different reasons for revisiting earlier work. Heer and Bostock repeated graphical perception experiments to assess the use of crowdsourcing for visualisation research~\cite{repvisCorpus02}. Kay and Heer reanalysed existing data to reconsider conclusions about correlation perception~\cite{repvisCorpus17}. Haehn et al.\ evaluated convolutional neural networks using graphical perception tasks previously studied with humans~\cite{repvisCorpus27}. These examples connect replication with evaluating research methods, reconsidering analyses, and examining how findings transfer to a different setting or subject.

\subsection{Describing and Designing Replication}

Terms such as strict and conceptual replication describe broad relationships to earlier work, while other terms, including triangulation, generalisation, and specification, draw attention to different strategies or aims~\cite{plesser2018_terminology,repvis1}. These need not be mutually exclusive: a study may revisit a finding while also examining its boundaries. More detailed representations already exist. Patil et al.\ provide a visual tool that separates components of the scientific process, including experimental design, data, and analysis, to compare study protocols and clarify definitions of reproducibility and replicability~\cite{patil2019_visual_tool}. Their approach shows how differences can be made visible without reducing a study to one overall label.

General experimental-design guidance supports choices about study tasks, procedures, participants, and analysis~\cite{purchase2012experimental,mackenzie2024human}. REPVIS2 addresses a different but related need: representing how those choices correspond to a particular reference study.

Earlier reviews have also examined replication practice directly. Hornb{\ae}k et al.\ investigated the extent and content of replications in HCI~\cite{hornbaek2014_once_enough}. In visualisation, Sukumar and Metoyer reviewed 16 replication studies and developed guidelines for making unbiased and meaningful replication design decisions~\cite{sukumar2018_unbiased}. In AR/VR, Arefin et al.'s review highlights that non-significant replication results may reflect limited sample sizes or violated statistical assumptions, underscoring the need to interpret outcomes alongside study design~\cite{arefin2025replication}. These studies provide an empirical basis for understanding how researchers revisit prior work. Our focus is on a representation through which the correspondence between a replication study and its reference study can be recorded consistently across design dimensions, supporting both comparison of published designs and consideration of future choices.

Our initial design space, REPVIS1, proposed such a dimension-based comparison~\cite{repvis1}. It treated the correspondence between studies as a set of design choices, supporting both retrospective characterisation and prospective planning. REPVIS1 also identified open questions concerning motivations, extensions, the granularity of its experiment dimension, and the comparison levels. REPVIS2 retains this perspective and examines these questions through application to published replication designs and evaluation beyond the cases that motivated the initial space.

\subsection{Developing Design Spaces}

Design spaces and typologies are established ways of organising visualisation research around comparable aspects of a task or design. Schulz et al.\ describe visualisation tasks through a multidimensional design space~\cite{schulz2013_task_design_space}. Brehmer and Munzner organise abstract tasks around why they are performed, how they are carried out, and what they operate on~\cite{brehmer2013_task_typology}. Munzner's nested model similarly relates design choices at different levels to corresponding validation concerns~\cite{munzner2009nested}. Such representations can therefore connect descriptions of existing work with decisions about work that has yet to be conducted.

Published literature can also inform the development of these structures. Lam et al.\ derived seven evaluation scenarios from a review of visualisation papers, providing guidance on evaluation goals and suitable approaches~\cite{lam2012_seven_scenarios}. Isenberg et al.\ subsequently used and extended this scheme in a systematic review~\cite{isenberg2013systematic}. This illustrates how a structured account of existing practice can help researchers plan future studies. We adopt this broader perspective for replication, where the object being described is the relationship between two study designs. Section~\ref{sec:method} explains how we used published studies to refine, evaluate, and apply the REPVIS2 design space. 

\section{Research Methodology}
\label{sec:method}

\begin{figure}[htbp]
    \centering
    \includegraphics[width=\columnwidth]{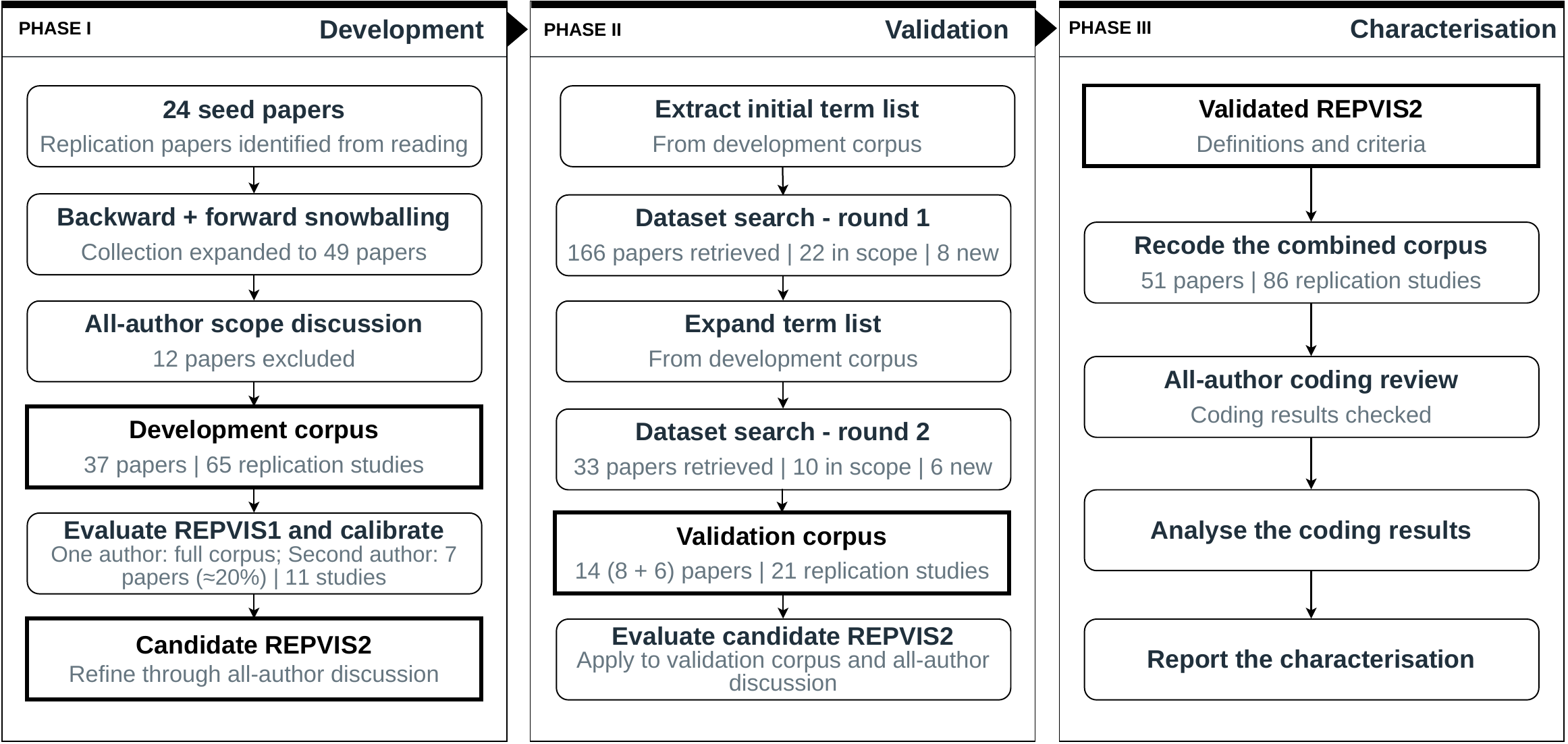}
    \caption{Development, validation, and characterisation workflow.}
    \label{fig:method-overview}
\end{figure}

We organised the work into development, validation, and characterisation (Figure~\ref{fig:method-overview}). We applied REPVIS1 to a development corpus, refined it into REPVIS2, froze its structure and criteria, and evaluated it on a separate validation corpus. We then recoded the combined corpus to characterise replication practices.

\subsection{Rationale}

A replication study does not have to repeat every part of a reference study in the same way. It may preserve one part of the design, change another to test generalisation, and add components to extend the original work. Such combinations are common in visualisation research. We therefore treat replication design as a multidimensional correspondence between a replication study and a reference study. Each dimension is compared separately, describing where the replication is close to its reference and where it deliberately departs. The resulting space is descriptive rather than prescriptive: it represents design possibilities without defining one design as a better or more valid form of replication.

Our design space applies this idea at the level of a replication--reference pair. The research question establishes the primary relationship between the studies. The dimensions describe how that relationship is implemented, avoiding reduction to one label or single type.

REPVIS1 was the first implementation of this view. It grew from our earlier work on replication terminology, advocacy, and practices in visualisation~\cite{repvis1}. It compared four practical dimensions: \textbf{\textsc{Experiment}}, \textbf{\textsc{Data}}, \textbf{\textsc{Participant}}, and \textbf{\textsc{Analysis}}, each coded as \textbf{\textit{identical}}, \textbf{\textit{similar}}, or \textbf{\textit{different}}.

REPVIS1 was a conceptual proposal illustrated through selected examples, not evaluated across broader replication practices. It remained unclear whether its four dimensions represented the design decisions found in practice, whether the three levels could be applied consistently, or whether essential design information remained outside the space.

We therefore treated REPVIS1 as a design space to be evaluated and refined through use. Our aim was not to capture every reported detail, but the information needed to describe how a replication design corresponded to, differed from, or extended its reference.

\subsection{Phase I: Development Corpus and REPVIS2 Development}

We consolidated 24 replication papers encountered during the development of REPVIS1. Backward and forward snowballing expanded this collection to 49 papers.

All authors reviewed the collection. We included papers reporting a study that revisited the same or a closely corresponding research question and compared findings with a reference study. Consensus discussion excluded 12 papers, leaving 37 in the development corpus.

Excluded papers typically reused an idea, task, stimulus, or method without revisiting the earlier question or comparing corresponding results. Kim and Heer used established findings to evaluate a new animation system~\cite{kim2021gemini}; Strain et al.\ reused scatterplot manipulations to investigate belief change rather than the earlier perceptual claim~\cite{strain2025scatterplot}. All authors discussed borderline cases until consensus.

One author applied REPVIS1 to the development corpus; a second independently coded approximately 20\% of the papers (7 papers, 11 studies) for calibration. All authors discussed disagreements and difficult cases to identify unclear definitions, overlapping concepts, and missing design distinctions. This produced a candidate REPVIS2 whose structure and criteria were frozen before validation. Section~\ref{sec:repvis2} reports the refinements.

\subsection{Phase II: Validation Corpus and Validation Procedure}

We next constructed a validation corpus containing papers that had not been used to develop REPVIS2. We searched the VisPubs dataset, a curated collection of publications from the IEEE VIS conference family, EuroVis, and visualisation-related CHI research~\cite{lange2024vispubs}. The dataset contained 6,253 papers published between 1986 and 2026, with titles and abstracts as searchable fields.

We used an iterative keyword search and applied the same scope criteria as for the development corpus. The first list was derived from the terminology found in the development corpus. It included variants of \textit{duplicate}, \textit{reevaluate}, \textit{rethink}, \textit{reassess}, \textit{revisit}, \textit{replicate}, and \textit{reproduce}. This search returned 166 papers and 22 papers were retained. Fourteen were already included in the development corpus, leaving eight new papers.

We checked whether the query recovered the development papers. Of the 27 indexed in VisPubs, 14 were found. Twelve of the 13 missed papers contained no replication-related terms in their title or abstract; their replication relationship became apparent only in the main text.

Following this audit, we searched with an expanded term list; the dataset had meanwhile been updated to include more recent papers. We added variants of \textit{follow-up experiment/study}, \textit{secondary analysis}, \textit{reanalysis}, and phrases combining \textit{previous} or \textit{prior} with \textit{finding} or \textit{result}. Excluding first-search results left 33 new papers. Ten were within scope; four overlapped with the development corpus, leaving six new papers. The eligible papers yielded no further title or abstract terms beyond the expanded query, so we ended the search. Terms and term-level results are provided in the supplementary material.

The largest term results came from \textit{reproduc*}, \textit{replicat*}, and \textit{revisit*}, which returned 65, 46, and 35 papers, with scope-positive rates of approximately 6.2\% (4/65), 41.3\% (19/46), and 14.3\% (5/35), respectively.

Together, the searches identified 32 in-scope papers (22+10), including 18 already in the development corpus. The remaining 14 formed the validation corpus and contained 21 replication studies.

We applied the frozen candidate to these 21 studies to evaluate design-information coverage, comparison-scale sufficiency, and structural stability. Section~\ref{sec:repvis2} defines these criteria and reports the result.

\subsection{Phase III: Corpus Coding and Characterisation}

After evaluating REPVIS2 on the validation corpus, we froze the final definitions and coding criteria. The first author then used these criteria to code the full development and validation corpora. The resulting codes were reviewed and discussed by all authors until consensus was reached.

The combined corpus contained 51 papers and 86 replication studies, with one coding row per study. Where a study drew on several predecessors, we selected one primary reference using the design anchor (the study that most directly shaped the design) and result anchor (the finding providing the main comparison). Supporting references did not create additional rows.

For each replication study, we identified and archived its reference report. Coding used paired reading of both reports, rather than relying only on the replication authors' account of the earlier work.

We summarised study-level counts and proportions for dimension codes, primary motivations, reported results, research-team relationships, explicit replication claims, and unified design choices. An interactive version of Figure~\ref{fig:replication-profiles} and the supplementary materials---including the search summary, corpus registry, REPVIS2 definitions and codebooks, and complete coding results---are available at \url{https://replication-study.github.io/}.

\section{Developing and Validating the REPVIS Design Space}
\label{sec:repvis2}

\begin{figure}[!t]
    \centering
    \includegraphics[width=\columnwidth]{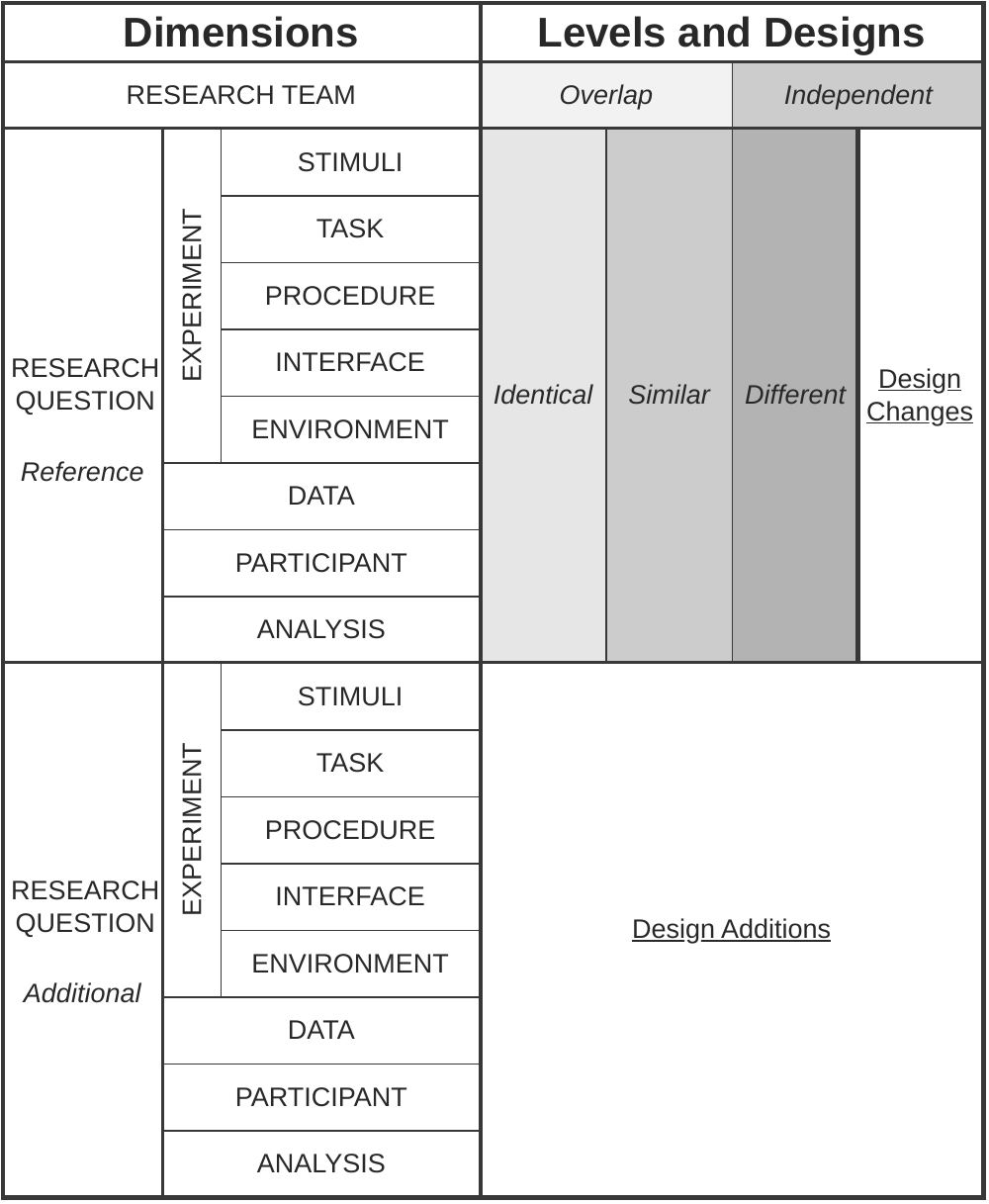}
    \caption{The REPVIS2 design space. Each of the eight dimensions can be applied to the study that replicates the reference study's research question, as well as to any additional study design that builds on the first.}
    \label{fig:repvis2-design-space}
\end{figure}

This section explains how development coding informed REPVIS2 and how the frozen candidate was evaluated on the 21 validation studies. Figure~\ref{fig:repvis2-design-space} presents the space, and Table~\ref{tab:repvis2-comparison-criteria} gives its operational criteria.

\subsection{Evaluation Criteria}

We evaluated REPVIS2 as a design space for describing the relationship between a replication study and its primary reference study.

We used three criteria: \textbf{coverage}, meaning that each replication--reference pair and its design-level information could be represented; \textbf{comparison-scale sufficiency}, meaning that observed relationships could be classified using the existing levels; and \textbf{structural stability}, meaning that no new dimension or structural change was required.

Development cases informed refinement; the validation corpus tested the frozen candidate on cases not used in its development.

\subsection{Refining the Dimensions}

REPVIS1 described replication design using four practical dimensions: \textbf{\textsc{Experiment}}, \textbf{\textsc{Data}}, \textbf{\textsc{Participant}}, and \textbf{\textsc{Analysis}}~\cite{repvis1}. The development coding showed that \textbf{\textsc{Experiment}} was too broad. A replication can preserve one part of an experiment while changing another. Assigning one comparison level to the whole experiment hid these mixed relationships and made it difficult to explain how the replication had been designed (as shown in Figure~\ref{fig:dimension-distribution}).

Several development corpus cases illustrated this limitation. Heer and Bostock retained a comparable graphical-judgement \textbf{\textsc{Task}} while adapting the \textbf{\textsc{Stimuli}}, \textbf{\textsc{Procedure}}, \textbf{\textsc{Interface}}, and \textbf{\textsc{Environment}} for Mechanical Turk~\cite{repvisCorpus02}. Kosara and Ziemkiewicz retained the \textbf{\textsc{Task}}, \textbf{\textsc{Procedure}}, and Java-based \textbf{\textsc{Interface}}, but moved from a supervised laboratory to remote participation~\cite{repvisCorpus04}. In their first bar-chart experiment, Talbot et al. retained the percentage-estimation \textbf{\textsc{Task}} while varying bar separation and distractors within the \textbf{\textsc{Stimuli}} to explain the reference finding~\cite{repvisCorpus14}. A single \textbf{\textsc{Experiment}} code would obscure these combinations of preservation and change, motivating separate comparison of its components.

REPVIS2 therefore replaces the \textbf{\textsc{Experiment}} dimension with five more specific dimensions. Together with \textbf{\textsc{Data}}, \textbf{\textsc{Participant}}, and \textbf{\textsc{Analysis}}, these form the eight practical dimensions:

\begingroup
\setlength{\parskip}{1.5pt}
\textbf{\textsc{Stimuli}} The object presented to participants for task execution.\par
\textbf{\textsc{Task}} The activity that participants are asked to execute using the stimuli.\par
\textbf{\textsc{Procedure}} The operational protocol through which the study is conducted.\par
\textbf{\textsc{Interface}} The system through which participants access the study.\par
\textbf{\textsc{Environment}} The surrounding setting where the study occurs.\par
\textbf{\textsc{Data}} The evidence source on which the study claims are ultimately based, such as answers, logs, and response time.\par
\textbf{\textsc{Participant}} The intended population and sampling approach, including recruitment and eligibility requirements.\par
\textbf{\textsc{Analysis}} The methods used to transform the collected evidence into claims.\par
\endgroup

We treat a \textbf{condition} as an experimental factor, not a dimension, and \textbf{context} as broad study background rather than a specific comparable component. \textbf{\textsc{Environment}} concerns the experimental setting, not its geographical location.

The corpus also contained human-to-machine replications, in which a machine or model performs the role of \textbf{\textsc{Participant}} in the reference study. We compare these cases using the same dimensions as other replications. These cases are highlighted in Figure~\ref{fig:replication-profiles}.

Although these eight dimensions are inter-related, they are defined and coded here as independently as possible. \textbf{\textsc{Research Team}} and \textbf{\textsc{Research Question}} are also important regarding the replication, but they play a different role. They describe the context of the replication relationship rather than a practical dimension of study implementation. \textbf{\textsc{Research Team}} records whether the replication and reference author teams are independent or overlap. \textbf{\textsc{Research Question}} separates the primary replication relationship from questions that extend beyond it. We therefore treat \textbf{\textsc{Research Team}} and \textbf{\textsc{Research Question}} as contextual dimensions and the other eight dimensions as practical dimensions.

\subsection{Selecting the Comparison Scale}

We considered a binary \textbf{\textit{same}}/\textbf{\textit{different}} scale, but it could not distinguish change that preserves a meaningful comparison from change beyond comparability. This distinction matters when replications test generalisation or introduce novelty.

We also explored finer distinctions based on set relationships. Published designs rarely supplied complete, well-bounded attribute sets, and changes within one dimension could imply conflicting relationships. Mathematical precision did not ensure clear semantic boundaries in actual designs.

We therefore refined the boundaries of three levels through development coding and calibration. The effort went into making these boundaries operationally clear, rather than introducing finer distinctions that were harder to apply. The validation cases required no additional level:

\textbf{\textit{Identical}} The dimension serves the same function and produces the same type of evidence as in the reference study. Unavoidable implementation differences may exist, but these are unintentional. Comparison is \textbf{direct} and like-for-like without extra interpretation.

\textbf{\textit{Similar}} The dimension differs substantively, so \textbf{direct} comparison is inappropriate. However, the dimension still plays a comparable function and yields a comparable type of evidence. So \textbf{comparability} is preserved.

\textbf{\textit{Different}} The dimension is so different that neither \textbf{direct} nor meaningful \textbf{comparability} is preserved, although the dimension is still important to the experiment design.

Table~\ref{tab:repvis2-comparison-criteria} gives dimension-specific criteria and examples. Two special rules address dimensions that are structurally absent in reanalysis or computational studies.

\noindent\textbf{Data reuse without recollection.} When a replication study reuses existing data rather than collecting new observations, it makes no new design choices about how that evidence was elicited or who produced it. \textbf{\textsc{Stimuli}}, \textbf{\textsc{Task}}, \textbf{\textsc{Procedure}}, \textbf{\textsc{Interface}}, \textbf{\textsc{Environment}}, and \textbf{\textsc{Participant}} therefore collapse. \textbf{\textsc{Data}} and \textbf{\textsc{Analysis}} remain active dimensions and are coded using their usual criteria.

\noindent\textbf{Computational studies.} When a computational replication study, such as a benchmark, simulation, or algorithm evaluation, has no participant role, \textbf{\textsc{Participant}} collapses. The other dimensions remain codable where they are instantiated by the study design.

\begin{table*}[t]
\caption{Comparison criteria and examples for the eight practical dimensions of REPVIS2. Dimensions are coded independently where possible.}
\label{tab:repvis2-comparison-criteria}
\centering
\scriptsize
\renewcommand{\arraystretch}{1.0}
\setlength{\tabcolsep}{2.5pt}
\newcommand{\criterionexample}[2]{\textbf{Criterion:} #1\par\textbf{Example:} #2}

\textbf{(a)} \textbf{\textsc{Stimuli}}, \textbf{\textsc{Task}}, \textbf{\textsc{Procedure}}, and \textbf{\textsc{Interface}}\par\vspace{1.5pt}
\begin{tabularx}{\textwidth}{@{}>{\raggedright\arraybackslash}p{0.064\textwidth}*{4}{>{\raggedright\arraybackslash}X}@{}}
\toprule
{} &
\textbf{\textsc{Stimuli}} &
\textbf{\textsc{Task}} &
\textbf{\textsc{Procedure}} &
\textbf{\textsc{Interface}} \\
\midrule
\textbf{\textit{Identical}} &
\criterionexample{Exactly the same stimuli are used; only unavoidable rendering or implementation differences exist.}{The same charts are shown with only minor display-resolution differences.} &
\criterionexample{Participants perform exactly the same task.}{Both studies ask which of two visual marks is larger.} &
\criterionexample{The same study protocol and experimental process are followed.}{Instructions, practice, timing, counterbalancing, and study structure are retained.} &
\criterionexample{The same device type, software, and input--output mechanisms are used.}{Both studies use the same static web interface with mouse input.} \\
\midrule
\textbf{\textit{Similar}} &
\criterionexample{The stimuli differ but serve a comparable perceptual, semantic, or experimental role.}{Both studies use pie charts with different values or scatterplots with different axes.} &
\criterionexample{Wording, difficulty, or operational details differ, but the task targets a comparable activity.}{'Choose the smaller value' becomes 'choose the larger value.'} &
\criterionexample{Some protocol elements change, but the core experimental structure remains comparable.}{The replication adds familiarisation, changes trial order, or changes from within- to between-participants.} &
\criterionexample{The interface changes, but the interaction logic remains comparable.}{A similar web implementation is used, or mouse input is replaced with touch for the same interaction.} \\
\midrule
\textbf{\textit{Different}} &
\criterionexample{The stimuli differ substantially in representation, form, or material.}{2D charts are replaced by immersive 3D physical objects.} &
\criterionexample{The tasks are not meaningfully comparable.}{Graph description replaces cluster identification.} &
\criterionexample{The experimental processes are substantially different.}{A controlled trial is replaced by a longitudinal field study.} &
\criterionexample{The interface changes the nature of interaction.}{A static desktop system is replaced by embodied VR interaction.} \\
\bottomrule
\end{tabularx}

\vspace{3pt}
\textbf{(b)} \textbf{ (continued) \textsc{Environment}}, \textbf{\textsc{Data}}, \textbf{\textsc{Participant}}, and \textbf{\textsc{Analysis}}\par\vspace{1.5pt}
\begin{tabularx}{\textwidth}{@{}>{\raggedright\arraybackslash}p{0.064\textwidth}*{4}{>{\raggedright\arraybackslash}X}@{}}
\toprule
{} &
\textbf{\textsc{Environment}} &
\textbf{\textsc{Data}} &
\textbf{\textsc{Participant}} &
\textbf{\textsc{Analysis}} \\
\midrule
\textbf{\textit{Identical}} &
\criterionexample{The setting and constraints are kept the same as far as possible.}{Both studies use the same quiet, supervised laboratory setting.} &
\criterionexample{Data form and values are the same.}{The exact same data values from the reference study are reused.} &
\criterionexample{The population and demographics are the same.}{Both studies recruit the same broad population with matching reported demographics.} &
\criterionexample{The same analysis approach is used.}{Both studies use the same statistical methods.} \\
\midrule
\textbf{\textit{Similar}} &
\criterionexample{The same setting type is retained, but constraints differ.}{Both studies take place in restaurants, but in different restaurants.} &
\criterionexample{Data form is the same, but the values differ.}{Both studies collect accuracy and response time from new observations.} &
\criterionexample{The population is the same, but demographics differ.}{Both recruit online adults but use different platforms or filters.} &
\criterionexample{The data are treated differently using a comparable approach.}{Regression replaces ANOVA to examine factor effects.} \\
\midrule
\textbf{\textit{Different}} &
\criterionexample{The setting is of a different type.}{A supervised laboratory study becomes an online study.} &
\criterionexample{The collected evidence is in a different form.}{Qualitative data replace quantitative data.} &
\criterionexample{The intention is to use a different population.}{University students are replaced by a target population of older adults.} &
\criterionexample{The analysis approach is substantively different.}{Video is analysed for quantitative metrics rather than qualitative themes.} \\
\bottomrule
\end{tabularx}
\end{table*}

\subsection{The Validated REPVIS2 Design Space}

Figure~\ref{fig:repvis2-design-space} summarises the final REPVIS2 design space. Each profile describes one replication--reference pair. \textbf{\textsc{Research Question}} contains two layers. The \textbf{\textit{reference}} layer is the same or closely corresponding question that establishes the replication relationship. The \textbf{\textit{additional}} layer contains questions pursued beyond that primary relationship.

Within the \textbf{\textit{reference}} \textbf{\textsc{Research Question}} layer, the eight practical dimensions describe how the replication design corresponds to the reference design. Each dimension is coded independently as \textbf{\textit{identical}}, \textbf{\textit{similar}}, or \textbf{\textit{different}}. This produces a profile of the experiment design. Two studies can therefore share the same broad motivation while occupying different positions in the design space, or use different implementations while retaining a comparable evidence chain.

Each replication can be expressed as a point in the design space:
\[
\begin{array}{@{}c@{}}
\mathrm{Replication}=(\mbox{\textsc{S}}_{r_1},\mbox{\textsc{T}}_{r_2},
\mbox{\textsc{Pr}}_{r_3},\mbox{\textsc{I}}_{r_4},\mbox{\textsc{E}}_{r_5},
\mbox{\textsc{D}}_{r_6},\mbox{\textsc{Pa}}_{r_7},\mbox{\textsc{A}}_{r_8}),\\[2pt]
r_i\in\{\mbox{\textit{identical}},\mbox{\textit{similar}},
\mbox{\textit{different}}\}
\end{array}
\]
For fully instantiated designs, the three levels yield \(3^8\) possible profiles, although some combinations may be impossible or inappropriate in a particular study context.

REPVIS2 also records design choices. \uline{Design Changes} describe modifications to the reference design within the primary replication relationship. \uline{Design Additions} describe components introduced beyond that relationship, including components used to pursue additional questions. These categories were not predefined. They emerged from the concrete design information extracted during coding and were subsequently unified separately for \uline{changes} and \uline{additions}, which are reported in Section~\ref{sec:result}. We kept them at a level that supports corpus-level frequency analysis without requiring a full thematic analysis. They can also be read as possible choices available to researchers when planning a future replication.

REPVIS2 represented all 21 validation studies and their design-level correspondence, \uline{changes}, and \uline{additions} without a new dimension, comparison level, or structural change. This supports its coverage and structural sufficiency for the validation corpus, without establishing that no future case could require refinement.

For corpus characterisation, the wider coding framework also records paper metadata, primary motivation, and reported replication result. These support interpretation but are not design dimensions, so they are not included in Figure~\ref{fig:repvis2-design-space} or the comparison criteria.

\section{Characterizing Replication Practices}
\label{sec:result}

We now use REPVIS2 to characterize the 51 papers and 86 replication studies in the combined corpus. We first describe when the studies were published and which earlier studies they revisited. We then examine why the replications were conducted, what results they reported, how their designs occupied the REPVIS2 space, and which \uline{design changes} and \uline{design additions} occurred repeatedly. Unless stated otherwise, the counts in this section refer to replication studies rather than papers.

Consider using REPVIS2 to characterise a replication to test whether a laboratory finding about visual metaphors holds with remote crowdworkers. Kosara and Ziemkiewicz's replication~\cite{repvisCorpus04} illustrates this design choice. The research teams overlapped. The replication retained the \textbf{\textsc{Stimuli}}, \textbf{\textsc{Task}}, \textbf{\textsc{Procedure}}, \textbf{\textsc{Interface}}, and \textbf{\textsc{Analysis}}, while changing the \textbf{\textsc{Environment}} and \textbf{\textsc{Participant}} population. Its profile is:
\[
\begin{array}{l}
\mathrm{Replication}=
(\mbox{\textsc{S}}_{\mathrm{I}},
 \mbox{\textsc{T}}_{\mathrm{I}},
 \mbox{\textsc{Pr}}_{\mathrm{I}},
 \mbox{\textsc{I}}_{\mathrm{I}},
 \mbox{\textsc{E}}_{\mathrm{D}},
 \mbox{\textsc{D}}_{\mathrm{S}},
 \mbox{\textsc{Pa}}_{\mathrm{D}},
 \mbox{\textsc{A}}_{\mathrm{I}}),\\[2pt]
\mathrm{I}=\mbox{\textit{identical}},\quad
\mathrm{S}=\mbox{\textit{similar}},\quad
\mathrm{D}=\mbox{\textit{different}}.
\end{array}
\]
New accuracy and response-time observations preserve the evidence form but change its values, making \textbf{\textsc{Data}} \textbf{\textit{similar}}. To plan a related replication, researchers can specify which correspondences to retain or vary, express these choices as a proposed profile, and document the concrete \uline{design changes} and their rationale. Components serving questions beyond the replication core are recorded as \uline{design additions} in the \textbf{\textit{additional}} \textbf{\textsc{Research Question}} layer.

\subsection{Corpus Overview}

\begin{figure}[tbp]
    \centering
    \includegraphics[width=\columnwidth]{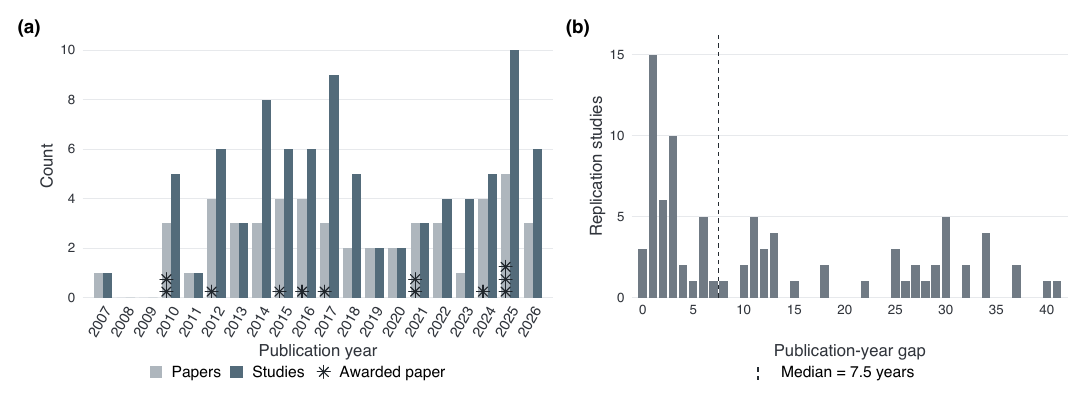}
    \caption{Publication timing of the corpus. (a) Papers and replication studies by year; stars mark awarded papers. (b) Publication-year gap between replication and primary reference studies.}
    \label{fig:corpus-metadata}
\end{figure}

The 51 papers were published between 2007 and 2026 (Figure~\ref{fig:corpus-metadata}a). The largest annual paper count was five in 2025. The largest study counts occurred in 2025 (10), 2017 (9), and 2014 (8). Twelve papers in the corpus received a best-paper or honorable-mention award, as marked by the stars. The 2026 count covers only part of the year. For journal papers associated with IEEE VIS family or EuroVis, year refers to the conference appearance year rather than the later journal issue year.

The time between a replication study and its primary reference study varied considerably (Figure~\ref{fig:corpus-metadata}b). Reference years follow the versions cited in the replication papers. The median gap was 7.5 years, and 46 of the 86 studies (53.5\%) appeared within ten years of their reference. The full range was zero to 41 years, with a mean of 12.8 years. The long right tail shows that the corpus contains both relatively immediate checks and much later returns to established findings.

The 86 studies referred to 48 distinct primary reference papers. Most were revisited only once: 33 were referenced by one replication study, and 44 by only one replication paper. Cleveland and McGill's 1984 \emph{Graphical Perception} study~\cite{cleveland1984graphical} served as the primary reference for 15 studies across eight papers. Heer and Bostock's 2010 \emph{Crowdsourcing Graphical Perception} study~\cite{repvisCorpus02} appeared in four studies across four papers, while Harrison et al.'s 2014 \emph{Ranking Visualizations of Correlation Using Weber's Law} study~\cite{repvisCorpus13} appeared in three studies across three papers. Bateman et al.'s 2010 \emph{Useful Junk?}~\cite{bateman2010usefuljunk} appeared in six studies, but these were concentrated in two papers. Thus, a small set of graphical-perception and perceptual-ranking studies acted as recurring cross-paper anchors, while the remaining references formed a long tail of more isolated replication activity.

\subsection{Motivations, Reporting, and Outcomes}

Across all 86 studies, 41 (47.7\%) reported results \emph{consistent} with their primary reference study, 30 (34.9\%) reported \emph{partly consistent} results, and 15 (17.4\%) reported results that were \emph{not consistent}. Note that the categories do not rank the quality or value of the studies; a result that is partly consistent or not consistent can be particularly informative when a study tests a boundary condition or transfers a finding to a new setting.

\begin{figure}[tbp]
    \centering
    \includegraphics[width=\columnwidth]{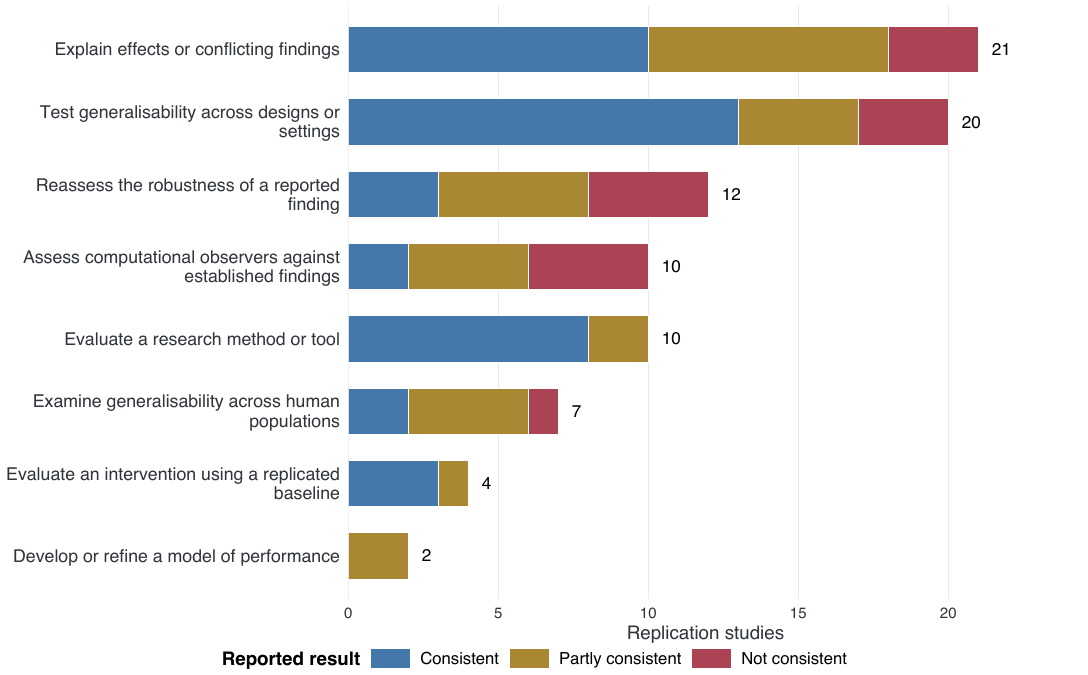}
    \caption{Primary motivations and reported replication results across the 86 studies.}
    \label{fig:motivation-outcomes}
\end{figure}

Each study was assigned one primary motivation. Figure~\ref{fig:motivation-outcomes} shows the frequency of these motivations together with the reported result. The most common motivations were to explain effects or conflicting findings (21 studies) and test generalisability across designs or settings (20). Twelve studies reassessed the robustness of a reported finding. Evaluating a research method or tool and assessing computational observers against established findings each accounted for ten. The remaining studies examined generalisability across human populations (7), evaluated an intervention using a replicated baseline (4), or developed or refined a model of performance (2).

The reported outcomes varied within most motivations. Studies explaining effects or conflicting findings included ten \emph{consistent}, eight \emph{partly consistent}, and three \emph{not consistent} results. Tests of generalisability across designs or settings included 13, four, and three, respectively. Robustness reassessments were more mixed, with three \emph{consistent}, five \emph{partly consistent}, and four \emph{not consistent} results. Studies assessing computational observers included two \emph{consistent}, four \emph{partly consistent}, and four \emph{not consistent} results. Some categories are small, so the patterns describe the present corpus rather than as evidence that transfer to computational participant generally produces results less consistent with prior findings.

\begin{figure}[tbp]
    \centering
    \includegraphics[width=\columnwidth]{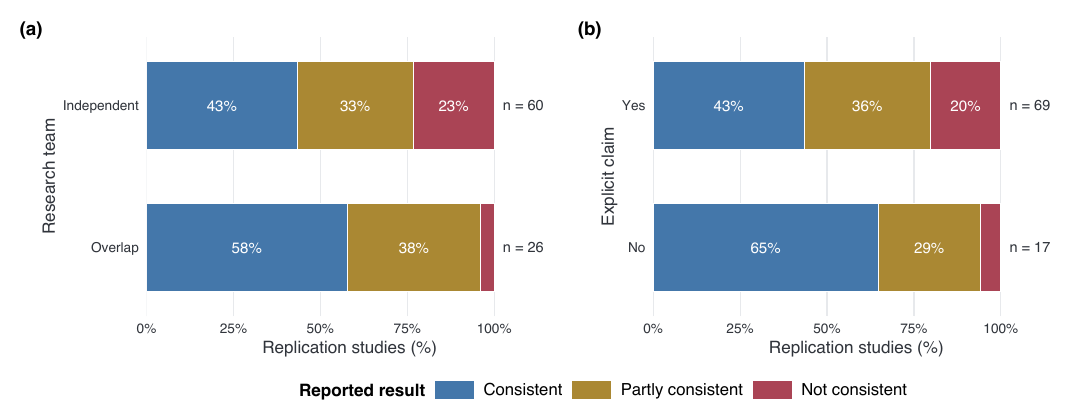}
    \caption{Reported replication results by research-team relationship and explicit replication claim. Percentages are normalized within each group; $n$ gives study counts.}
    \label{fig:team-claim-outcomes}
\end{figure}

Figure~\ref{fig:team-claim-outcomes} relates the reported outcomes to the two contextual reporting fields. Sixty studies were conducted by independent research teams, while 26 had at least one author overlapping with the reference study. \emph{Consistent} results accounted for 43.3\% of the independent-team studies and 57.7\% of the overlapping-team studies; \emph{not consistent} results accounted for 23.3\% and 3.8\%, respectively. An explicit replication claim was present for 69 studies and absent for 17. Among the explicit claims, 43.5\% reported \emph{consistent} results and 20.3\% \emph{not consistent} results. Among the studies without an explicit claim, 64.7\% reported \emph{consistent} results and 5.9\% \emph{not consistent} results. These are descriptive differences within the corpus. Team composition and explicit wording are entangled with topic, design, publication, and reporting choices, so the figure does not support a causal interpretation. In particular, the absence of an explicit claim does not mean that a study is not a replication under our scope definition (whether it is a replication study) because we determine the scope by the reported information in the paper.

\subsection{Design Space Profiles}

\begin{figure}[tbp]
    \centering
    \includegraphics[width=\columnwidth]{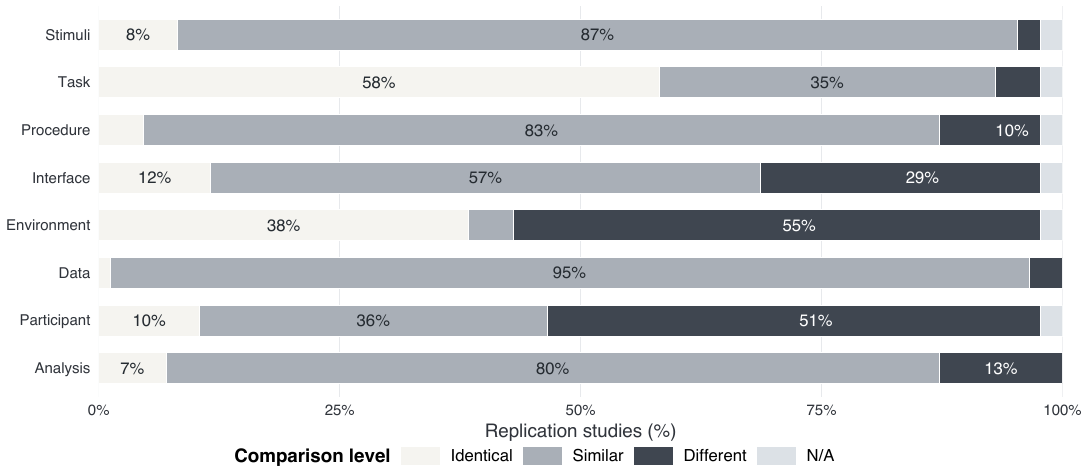}
    \caption{Distribution of comparison levels across the eight practical dimensions. Percentages are calculated within each dimension.}
    \label{fig:dimension-distribution}
\end{figure}

Figure~\ref{fig:dimension-distribution} summarizes the comparison levels across the eight practical dimensions. Five dimensions were most often coded as \textbf{\textit{similar}}: \textbf{\textsc{Stimuli}} (75 of 86), \textbf{\textsc{Procedure}} (71), \textbf{\textsc{Interface}} (49), \textbf{\textsc{Data}} (82), and \textbf{\textsc{Analysis}} (69). \textbf{\textsc{Interface}} also had 25 studies coded as \textbf{\textit{different}}. \textbf{\textsc{Task}} was the only dimension for which a majority of studies were \textbf{\textit{identical}} (50). By contrast, \textbf{\textsc{Environment}} and \textbf{\textsc{Participant}} were most often \textbf{\textit{different}}, in 47 and 44 studies, respectively.

Because \textbf{\textit{identical}} \textbf{\textsc{Data}} means reuse of the same evidence values, an all-\textbf{\textit{identical}} profile is not the appropriate baseline for a newly conducted replication. The closest newly collected profile would instead have \textbf{\textit{similar}} \textbf{\textsc{Data}} and \textbf{\textit{identical}} codes on the other seven dimensions. No study occupied this profile. The closest fully instantiated design had \textbf{\textit{similar}} \textbf{\textsc{Stimuli}} and \textbf{\textsc{Data}}, with the other six dimensions coded as \textbf{\textit{identical}}~\cite{repvisCorpus05}. Thus, every newly conducted study made at least one additional design change beyond recollecting the same form of evidence. More broadly, the distribution suggests a recurring design pattern. Replications often retained the participant activity while adapting the operational shell around it. The \textbf{\textsc{Task}} was frequently retained, the \textbf{\textsc{Stimuli}}, \textbf{\textsc{Procedure}}, \textbf{\textsc{Interface}}, \textbf{\textsc{Data}}, and \textbf{\textsc{Analysis}} usually remained comparable, and the \textbf{\textsc{Environment}} and \textbf{\textsc{Participant}} often changed more substantially.

The parallel-coordinate view in Figure~\ref{fig:replication-profiles} retains each complete coding row. It connects the source paper and research-team relation to the eight practical dimensions, with color showing the reported result. The 86 studies occupied 44 distinct eight-dimensional profiles. Profiles retain collapsed dimensions as recorded. This diversity would be hidden by a single label such as strict, partial or conceptual replication. At the same time, the repeated paths through the figure make the dominant pattern visible: an \textbf{\textit{identical}} \textbf{\textsc{Task}} often co-occurs with \textbf{\textit{similar}} operational and evidential dimensions and \textbf{\textit{different}} \textbf{\textsc{Environment}} or \textbf{\textsc{Participant}}.

Studies in which a machine or model performs the participant role connect to the \emph{Machine} extension below the \textbf{\textit{different}} bar on the \textbf{\textsc{Participant}} axis. This highlights machine-based replications within the existing \textbf{\textit{different}} category.

We also summarized recurring combinations of design departures. For the seven non-\textbf{\textsc{Data}} dimensions, both \textbf{\textit{similar}} and \textbf{\textit{different}} counted as a departure from \textbf{\textit{identical}}. For \textbf{\textsc{Data}}, however, \textbf{\textit{similar}} was treated as the expected baseline for newly collected same-form evidence and only \textbf{\textit{different}} counted as an additional departure. Collapsed dimensions did not count as departures. Under this definition, the most common combination covered every non-\textbf{\textsc{Data}} dimension except \textbf{\textsc{Task}} (26 studies, 30.2\%): nine results were \emph{consistent}, 11 \emph{partly consistent}, and six \emph{not consistent}. The second covered all seven non-\textbf{\textsc{Data}} dimensions (14, 16.3\%; six \emph{consistent}, four \emph{partly consistent}, and four \emph{not consistent}). The third covered every non-\textbf{\textsc{Data}} dimension except \textbf{\textsc{Environment}} (10, 11.6\%; six \emph{consistent}, two \emph{partly consistent}, and two \emph{not consistent}). Together, these three combinations accounted for 50 of the 86 studies (58.1\%). This summary records where an additional design departure occurred, not its severity, and the mixed outcomes within each combination do not suggest a simple profile--result relation. Figure~\ref{fig:replication-profiles} retains the complete level assignment, which is essential for interpreting the profiles.

\subsection{Recurring Design Choices}

The unified design choices make the concrete adaptations behind these profiles available for frequency analysis. Tables~\ref{tab:unified-design-changes} and~\ref{tab:unified-design-additions} report 22 \uline{change} phrases and 21 \uline{addition} phrases, derived separately from the respective descriptions. For these summaries, the five closely related experimental dimensions are grouped under \textbf{\textsc{Experiment}} to keep connected design choices together. Each phrase is counted at most once per replication study. A study can contain several choices, so the counts are not mutually exclusive and do not sum to 86. An additional design shared by several replication studies contributes to each study to which it applies. These frequencies describe the presence of choices across replication designs.

\begin{table}[t]
\caption{Unified \uline{design changes} counted at study level ($N=86$).}
\centering
\footnotesize
\setlength{\tabcolsep}{3pt}
\renewcommand{\arraystretch}{1.08}
\begin{tabularx}{\columnwidth}{@{}>{\raggedright\arraybackslash}Xr@{}}
\toprule
\textbf{Unified phrase} & \textbf{$n$} \\
\midrule
\multicolumn{2}{@{}l}{\textbf{\textsc{Experiment}}} \\
Change the study interface & 72 \\
Adjust the study administration protocol & 70 \\
Reconstruct or adapt stimulus materials & 47 \\
Change the study setting & 46 \\
Restrict stimulus or condition coverage & 33 \\
Change the task or response requirement & 33 \\
Change the experimental structure or paradigm & 32 \\
Expand stimulus or condition coverage & 26 \\
Change stimulus representation or modality & 19 \\
\addlinespace[2pt]
\multicolumn{2}{@{}l}{\textbf{\textsc{Data}}} \\
Collect new observations & 84 \\
Revise the evidence measures collected & 38 \\
Reuse existing evidence & 5 \\
Replace the primary evidence type & 3 \\
\addlinespace[2pt]
\multicolumn{2}{@{}l}{\textbf{\textsc{Participant}}} \\
Adjust sampling and recruitment & 45 \\
Change the target human population & 34 \\
Replace the participant agent & 10 \\
\addlinespace[2pt]
\multicolumn{2}{@{}l}{\textbf{\textsc{Analysis}}} \\
Change statistical tests or models & 57 \\
Revise uncertainty or robustness assessment & 48 \\
Change the scope of comparisons & 42 \\
Change outcome scoring or aggregation & 36 \\
Analyse individual differences or moderators & 24 \\
Change qualitative or visual analysis & 6 \\
\bottomrule
\end{tabularx}
\label{tab:unified-design-changes}
\end{table}

\begin{table}[t]
\caption{Unified \uline{design additions} counted at study level ($N=86$).}
\centering
\footnotesize
\setlength{\tabcolsep}{3pt}
\renewcommand{\arraystretch}{1.08}
\begin{tabularx}{\columnwidth}{@{}>{\raggedright\arraybackslash}Xr@{}}
\toprule
\textbf{Unified phrase} & \textbf{$n$} \\
\midrule
\multicolumn{2}{@{}l}{\textbf{\textsc{Experiment}}} \\
Add a condition or benchmark comparison & 14 \\
Add a method or model evaluation & 10 \\
Add a new task or question & 8 \\
Add a formative or pilot study & 4 \\
Add a qualitative inquiry component & 3 \\
Add a follow-up assessment or questionnaire & 3 \\
\addlinespace[2pt]
\multicolumn{2}{@{}l}{\textbf{\textsc{Data}}} \\
Collect additional performance evidence & 29 \\
Collect participant-background or ability measures & 24 \\
Collect qualitative accounts or recordings & 23 \\
Record additional operational or contextual metadata & 15 \\
Collect subjective ratings or preferences & 13 \\
Record additional process or interaction evidence & 8 \\
Derive additional features or evidence representations & 4 \\
\addlinespace[2pt]
\multicolumn{2}{@{}l}{\textbf{\textsc{Participant}}} \\
Recruit a separate extension sample & 13 \\
Recontact or select existing participants & 3 \\
\addlinespace[2pt]
\multicolumn{2}{@{}l}{\textbf{\textsc{Analysis}}} \\
Analyse additional outcomes or comparisons & 23 \\
Analyse background or contextual associations & 16 \\
Analyse additional qualitative accounts & 15 \\
Develop or evaluate additional models & 8 \\
Synthesise findings across studies & 7 \\
Assess operational feasibility or study process & 6 \\
\bottomrule
\end{tabularx}
\label{tab:unified-design-additions}
\end{table}

Table~\ref{tab:unified-design-changes} shows \uline{changes} within the replication relationship. The most frequent choices were collecting new observations (84 studies), changing the study interface (72), and adjusting the study administration protocol (70). Changing statistical tests or models (57) and revising uncertainty or robustness assessment (48) were also common. Recruitment and sampling were adjusted in 45 studies, while 34 changed the target human population. These counts give a more concrete account of the dimensional pattern above: many replications retained a comparable question while adapting how the study was delivered, sampled, measured, and analysed.

Five studies reused existing evidence: two relied on reuse without collecting new observations, while three combined reused evidence with new observations. These groups therefore overlap with the 84 studies collecting new observations. Reuse does not automatically imply \textbf{\textit{identical}} \textbf{\textsc{Data}}: one of these five studies was coded as \textbf{\textit{identical}} and four as \textbf{\textit{similar}} because the author deliberately added or removed the values that the reference collected but did not use. Likewise, collecting new observations is a design choice recorded in the table, not by itself a change in evidence type.

\uline{Additions} occurred in 63 studies (73.3\%). Table~\ref{tab:unified-design-additions} shows that they were not limited to extra conditions or experiments. Collecting additional performance evidence was most frequent (29 studies), followed by participant-background or ability measures (24) and qualitative accounts or recordings (23). Analysing additional outcomes or comparisons also occurred in 23 studies. Fourteen added a condition or benchmark comparison, and 13 recruited a separate extension sample. These \uline{additions} provided further evidence about performance, participants, and context alongside the primary replication. The four studies with a formative or pilot \uline{design additions} record it as a supporting design.

Taken together, these two tables can also be used to inspire future replication study design as possible design choices. The results also show replication practice as a combination of preservation, adaptation, and extension. The corpus contains few repeated reference anchors but a wide range of design profiles around them. Replications commonly preserve a comparable task and evidence chain while changing recruitment, setting, procedure, interface, and analysis. They also use the replicated relationship as a platform for additional conditions, measures, and analyses. REPVIS2 keeps these decisions visible.

\section{Discussion}
\label{sec:discussion}

\subsection{RQ1: Representing Replication Designs}

\textbf{Dimensional correspondence makes replication design explicit.} REPVIS2 represents a replication through its correspondence with a reference study across eight practical dimensions. Separate comparison levels make mixed relationships visible: a study can retain its \textbf{\textsc{Task}} while changing its \textbf{\textsc{Procedure}} and \textbf{\textsc{Environment}} to different degrees. \textbf{\textsc{Research Team}} and \textbf{\textsc{Research Question}} describe the context of this comparison, while \uline{design changes} and \uline{design additions} record the concrete choices within and beyond the replication core. Motivations and reported results supplement this representation in the wider coding framework. The 21 validation studies required no additional dimension, comparison level, or structural change, supporting the design space's coverage of these cases.

\textbf{Design profiles go beyond overall replication labels.} This representation develops the perspective proposed in REPVIS1~\cite{repvis1}. REPVIS2 complements the approaches reviewed in Section~\ref{sec:related-work}. High-level terms such as conceptual replication, generalisation, and triangulation communicate broad strategies but do not specify a unique design profile~\cite{plesser2018_terminology,repvis1}. Generalisation may involve changing \textbf{\textsc{Participant}}, \textbf{\textsc{Environment}}, or both. Reanalysis can instead leave only \textbf{\textsc{Data}} and \textbf{\textsc{Analysis}} active. REPVIS2 makes these distinctions explicit while allowing familiar terms to remain useful.

\textbf{Explicit comparisons connect individual designs into a profile.} Other representations also distinguish study components. Patil et al.\ separate population, experimental design, data, and analysis, and distinguish differences from missing or incorrectly reported information~\cite{patil2019_visual_tool}. Arefin et al.\ compare selected attributes of AR/VR replications, including experimental design, participant characteristics, and displays~\cite{arefin2025replication}. REPVIS2 organises the comparison through eight dimensions with explicit criteria for \textbf{\textit{identical}}, \textbf{\textit{similar}}, and \textbf{\textit{different}}. In particular, \textbf{\textit{similar}} captures change that preserves a meaningful basis for comparison. The assignments form a complete profile, and the two research question layers distinguish changes within the replication core from additional inquiry. Task typologies and replication guidelines can help researchers specify and justify their choices~\cite{brehmer2013_task_typology,sukumar2018_unbiased}. REPVIS2 then records how the resulting study relates to its reference. It provides neither a substitute for those detailed accounts nor a checklist that guarantees a sound study. Its contribution is a precise structure for systematically describing these decisions.

\subsection{RQ2: Patterns in Replication Practice}

\textbf{The corpus shows diverse but recurring forms of adaptation.} Its 86 studies occupied 44 profiles, with a dominant pattern of retaining the activity under investigation while adapting the surrounding study. \textbf{\textsc{Task}} was often \textbf{\textit{identical}}; several operational and evidential dimensions were commonly \textbf{\textit{similar}}; and \textbf{\textsc{Environment}} and \textbf{\textsc{Participant}} were more often \textbf{\textit{different}}. These combinations show why comparison at the level of individual dimensions is useful. Design variation can also clarify earlier findings: Purchase shows how changes in stimulus presentation can expose unintended methodological biases~\cite{purchase2014healthy}.

\textbf{Replication serves purposes beyond pure validation.} The motivations and \uline{additions} show that studies sought explanations for effects or conflicting findings, tested generalisability and robustness, evaluated methods and computational observers, and used replicated baselines to assess interventions. \uline{Additions} supplied further performance, participant-background, and qualitative evidence. Results that are \emph{partly consistent} or \emph{not consistent} can therefore be informative when studies examine boundaries or transfer. Mixed outcomes within common motivations and profiles do not support a simple design--result relationship. Differences associated with team overlap and explicit replication claims may reflect tacit knowledge, topic selection, design, or reporting; these influences cannot be separated here.

\textbf{Participant variation as complexity.} The \textbf{\textsc{Participant}} distribution also reflects the complexity of visualisation research involving human subjects. Differences in age, expertise, accessibility, language, culture, and recruitment affect who provides evidence and to whom findings may apply. Participant variation is therefore part of the interpretation of a replication, as well as a practical recruitment decision.

\textbf{The reference distribution reveals an uneven pattern of follow-up within the corpus.} A small set of graphical-perception and perceptual-ranking studies served as recurring references across papers, while most primary references appeared in only one replication paper. The corpus thus contains both repeated investigations of established findings and more isolated returns to earlier work.

\subsection{RQ3: Implications for Future Planning and Reporting}

\textbf{Characterisation can inform prospective planning.} REPVIS1 proposed retrospective characterisation and prospective planning as two uses of the design space~\cite{repvis1}. The present work demonstrates characterisation at corpus scale and provides concrete design choices that researchers can consider when planning a replication. Starting from a reference study and claim, researchers can specify which correspondences to preserve and which to vary. The choices in Tables~\ref{tab:unified-design-changes} and~\ref{tab:unified-design-additions} provide options for adapting materials, recruitment, evidence collection, and analysis. Recording a proposed profile and the rationale for its \uline{changes} and \uline{additions} makes these decisions available for comparison before data collection.

\textbf{Observed patterns can inform design priorities.} Common profiles provide precedents; less frequent combinations can suggest questions where the study context makes them meaningful. For example, a \textbf{\textit{different}} \textbf{\textsc{Task}} or \textbf{\textsc{Data}} may support triangulation. Related work on contrasting future scenarios illustrates how variation can reveal shared and context-dependent findings~\cite{salovaara2025triangulating}. Because several dimensions often change together, we cannot isolate the effects of individual choices. The corpus can, however, motivate replications that vary selected dimensions while retaining an interpretable relationship to the reference. Low frequency alone does not establish a useful research gap, since some combinations may be inappropriate or impossible. Priorities can also reflect a claim's importance, uncertainty, downstream use, and the consequences of relying on it if it is wrong, extending cumulative investigation beyond the corpus's recurring reference studies.

\textbf{Participant variation as opportunity.} Participant variation creates opportunities for testing generalisability and transfer. For example, researchers can examine whether findings apply to children, older adults, domain experts, or people with different access needs. The movement from human participants to machine actors is an emerging extension of this opportunity. Models taking the participant role extend these questions to human-to-machine replication~\cite{repvisCorpus27,repvisCorpus50}. These cases show both that the current dimensions can extend beyond human-subject studies and that future machine, simulation, and embodied replications may create further room for adjustment. The mixed results for computational observers suggest that correspondence with established human study findings should be tested in machines rather than assumed. Future replications can compare models with different populations, identify which tasks transfer, and examine sensitivity to model version, prompting, sampling, and run protocol. These details span several dimensions and should be reported independently.

\textbf{Making replication relationships reportable.} Explicit reporting makes replication relationships inspectable. The reporting difficulties encountered during corpus construction point to a further use. Twelve of the 13 development papers missed by the first VisPubs query contained no replication-related term in their title or abstract, and 17 coded studies made no explicit replication claim. Uneven reporting of references, populations, protocols, and evidence relationships also complicated comparison, consistent with related meta-research in visualisation~\cite{isenberg2013systematic}. A design space cannot recover details that were not reported. For future studies, REPVIS2 can provide a lightweight structure for clearer reporting: identify the primary reference and corresponding question; state the profile, including collapsed dimensions; explain the \uline{design changes} and \uline{design additions}; and identify the findings supporting the reported result. Signalling replication in titles, abstracts, or keywords would improve discovery. Clearer records would also support later analyses accounting for topic, reference study, and studies nested within papers.

\textbf{Transfer beyond visualisation.} Other technology-oriented human-subject fields share similar experimental design practices. REPVIS2 may help structure replication relationships in these fields, although its transferability requires evaluation against their own research practices.

\subsection{Future Work}

\textbf{The corpus is constructed rather than exhaustive.} We constructed the corpus using complementary discovery strategies because replication relationships are not consistently signalled in titles or abstracts. Development corpus papers came from prior reading and snowballing, while validation papers were found through title and abstract searches. This source covers selected venues, and papers that do not signal replication in searchable fields are less likely to be found. The reported distributions therefore characterise the identified corpus rather than the exhaustive prevalence of replication across all visualisation venues. Future work could extend venue coverage, search full text and citation networks, and update the corpus over time.

\textbf{Scope, pairing, and coding required further judgement.} All authors discussed borderline papers, but screening was not pre-registered and did not retain a complete structured exclusion log. Selecting one primary reference using design and result anchors also simplifies studies that synthesise several prior works. The full corpus was coded by the first author using frozen consensus criteria. A second author independently coded 20\% of the development papers for calibration, and all authors discussed the final decisions, but no formal inter-rater reliability statistic was calculated. Future work could retain secondary references as a network and test the criteria through external independent coding.

\textbf{The analysis is limited by published reporting and by its descriptive coding.} Population intention, protocol details, evidence forms, and analysis choices were not equally available. The \emph{consistent}, \emph{partly consistent}, and \emph{not consistent} categories summarise author-reported results, rather than standardised effect estimates or study quality. One primary motivation suppresses secondary motives, and the unified choices are a lightweight consolidation rather than an exhaustive thematic taxonomy. These choices support mapping the corpus, but not causal conclusions about profiles, teams, explicit claims, and outcomes. Also, the collected papers may be subject to publication bias, as the requirement for novelty means that ``pure'' replications, or replications that introduce little or nothing new, are often perceived as lacking sufficient novelty for publication.

\textbf{The validation process is constrained by the existing replication practices.} Finally, the validation corpus contained 21 studies. It supports coverage and structural stability, but cannot guarantee that no future replication will require refinement. Emerging human-machine studies may create further boundary cases. The prospective planning function has also not been evaluated in practice. Future work could use REPVIS2 during real replication planning and examine whether it improves design justification, comparison, and reporting consistency. 

\section{Conclusion}
\label{sec:conclusion}

Replication cannot be adequately described by a single study-type or outcome label. REPVIS2 makes its design explicit through eight practical dimensions and three comparison levels, alongside research question and research team context. Application to a separate validation corpus supports its coverage and structural sufficiency.

Applying REPVIS2 to 51 papers and 86 replication studies showed that visualisation replication practice is diverse but not unstructured. The studies occupied 44 profiles, often retaining a comparable task while adapting procedure, interface, environment, participant population, evidence, or analysis, and many used replication as a basis for further measures, conditions, comparisons, and analyses. Reported outcomes remained mixed across common profiles. These patterns describe how replication has been practiced; they do not establish one preferred design or a simple relationship between design choices and reported replication results.

REPVIS2 supports characterising and reporting existing studies and planning future replications. Researchers can specify the correspondence needed for a claim, consider design choices, and communicate their relationship to prior work. It treats replication not as a simple copy, but as a design that can be compared, described, and planned.



\bibliographystyle{abbrv-doi-hyperref}

\bibliography{template,corpus}

\appendix 
\crefalias{section}{appendix} 

\end{document}